\documentclass[twocolumn]{aastex631}
\usepackage{booktabs}
\usepackage{amsmath}
\usepackage{amsfonts}
\usepackage{graphicx}
\usepackage{xspace}
\usepackage{soul}
\usepackage{xcolor}
\renewcommand{\hl}[1]{#1}

\newcommand{\aestra}{\texttt{AESTRA}\xspace}
\newcommand{\metersec}{$\rm{m\,s}^{-1}$~}

\soulregister{\aestra}{7}
\soulregister{\citep}{7}
\soulregister{\cite}{7}
\soulregister{\autoref}{7}

\begin{document}

\title{AESTRA: Deep Learning for Precise Radial Velocity Estimation in the Presence of Stellar Activity}

\author[0000-0002-1001-1235]{Yan Liang}
\affiliation{Department of Astrophysical Sciences, Princeton University,
Princeton, NJ 08544, USA}
\author[0000-0002-4265-047X]{Joshua N.\ Winn}
\affiliation{Department of Astrophysical Sciences, Princeton University,
Princeton, NJ 08544, USA}
\author[0000-0002-8873-5065]{Peter Melchior}
\affiliation{Department of Astrophysical Sciences, Princeton University, Princeton, NJ 08544, USA}
\affiliation{Center for Statistics \& Machine Learning, Princeton University, Princeton, NJ 08544, USA}


\begin{abstract}

Stellar activity interferes with precise radial velocity measurements and limits our ability to detect and characterize planets, particularly Earth-like planets. We introduce \aestra (Auto-Encoding STellar Radial-velocity and Activity), a deep learning method for precise radial velocity measurements. It combines a spectrum
auto-encoder, which learns to create realistic models of the star's rest-frame spectrum, and a radial-velocity estimator, which learns to identify true Doppler shifts in the presence of spurious shifts due to line-profile variations.
Being self-supervised, \aestra does not need ``ground truth'' radial velocities
for training, making it applicable to exoplanet host stars for which the truth
is unknown. In tests involving 1{,}000 simulated spectra, \aestra can detect planetary signals as low as 0.1\,\metersec even in the presence of 3\,\metersec of activity-induced noise and 0.3\,\metersec of photon noise per spectrum.
\end{abstract}



\section{Introduction} \label{sec:intro}

The search for Earth-like planets in the habitable zones of Sun-like stars is one of the premier challenges of modern astronomy. 
Despite significant advancements in instrumentation and data analysis techniques, such as the achievement of $\sim$0.3~\metersec precision in recent ESPRESSO observations of Proxima Centauri \citep{mascareno2020revisiting}, the detection of Earth-like planets will require further advances. 
The next generation of Extremely Precise Radial Velocity (EPRV) instruments is designed to achieve an instrumental precision of $\sim$0.1~\metersec \citep{blackman2020performance,hall2018feasibility}. However, the emergent spectrum of a star is inherently variable, leading to
apparent variations in the star's radial velocity that can mimic or mask planetary signals.
This ``radial-velocity jitter'' arises from magnetic activity, granulation, and acoustic oscillations, which have different characteristic
amplitudes and timescales \citep{haywood2014planets,mascareno2020revisiting,milbourne2019harps}.
Given that even the quietest stars exhibit RV jitter of approximately 2~\metersec \citep{brems2019radial}, it is essential to mitigate the influence of stellar activity signals on EPRV measurements if we are to have any chance of detecting Earth-like exoplanets.

Numerous data analysis techniques have been proposed to determine true RVs from stellar spectra in the presence of activity.
A common approach is to extract RVs using traditional techniques, such as cross-correlation with a spectral template,
and then de-correlate the apparent RV time series with various activity indices that are derived independently
from the spectra \citep{milbourne2019harps,giguere2016combined}. Another approach, sometimes combined with the previous
approach, is to model the RV time series
using Gaussian Process (GP) regression to represent the effects of activity \citep{haywood2014planets,mascareno2020revisiting}. 
There are also data-driven techniques, such as performing shift-invariant operations on the cross-correlation function (CCF) to isolate activity-induced signals \citep{collier2021separating}, and performing full spectral modeling to address contamination due to variability in telluric absorption lines \citep{bedell2019wobble}. 
Supervised machine learning techniques have been applied to recognize spurious RV signals associated with
changes in shape of the CCF \citep{de2022identifying}.

On the observational side, there has been a growing trend toward more intensive and longer-term monitoring of stars. It is no
longer unusual to see hundreds of RVs gathered for a single object, and we will probably
begin seeing thousands of RVs in the coming decade \citep{crass2021extreme}.
For example, the Terra Hunting Experiment will use a high-resolution echelle spectrograph and an automated
2.5\,m telescope to measure RVs of a few dozen stars on every clear night for at least 10 years \citep{thompson2016harps3,hall2018feasibility}.  Such intensive monitoring may open up new possibilities for
stellar activity mitigation by providing larger samples of activity-related perturbations.


In this paper, we present \aestra\ (Auto-Encoding STellar Radial-velocity and Activity), a deep learning method for distinguishing between true Doppler shifts and activity-induced spectral perturbations, given a large sample of stellar spectra covering a variety of the star's activity states and Doppler shifts.
The architecture of \aestra combines a neural network for radial-velocity estimation, and a spectrum auto-encoder 
called \texttt{spender} \citep{melchior2023autoencoding,liang2023autoencoding} for activity modeling. 
\aestra is designed to disentangle pure Doppler shifts from apparent Doppler shifts due to spectral perturbations,
without any prior knowledge of the star's spectrum or orbital motion.

The spectrum auto-encoder architecture \texttt{spender} was originally built for analyzing and representing galaxy spectra in the presence of differing
cosmological redshifts. It features an attentive autoencoder
that scans over the wavelength range and compresses the information
in each spectrum into a small number of parameters,
which are ideally independent of redshift \citep{liang2023autoencoding}, 
and a decoder capable of transforming the parameters
back into an accurate reconstruction of the spectrum 
in the rest frame (i.e., without any redshift).
\texttt{Spender} proved capable
of compressing $\sim$500{,}000 galaxy spectra from the Sloan Digital Sky Survey \citep{ahumada202016th} into a low-dimensional latent space while maintaining high fidelity in the reconstructed spectra 
\citep{melchior2023autoencoding,liang2023autoencoding}.
Cosmological redshifts are many orders of magnitude larger than the Doppler shifts associated with exoplanets; for a Earth-like planet around a
Sun-like star, $z\sim 10^{-10}$.
Nevertheless, we wondered whether the ability of \texttt{spender} to identify and isolate intrinsic spectral features while
disregarding Doppler shifts and other artifacts would be applicable to exoplanet detection.

The remainder of the paper is structured as follows. \autoref{sec:method} describes the architecture
of \aestra and the loss functions that are used for training. \autoref{sec:application} presents the application
of \aestra to simulated datasets. Finally, \autoref{sec:conclude} summarizes our findings,
discusses limitations and
potential extensions of \aestra,
and discusses potential applications to solar spectroscopy and exoplanet surveys.



\section{Methods} \label{sec:method}

\begin{figure*}[t!]
\centering
\includegraphics[width=\textwidth]{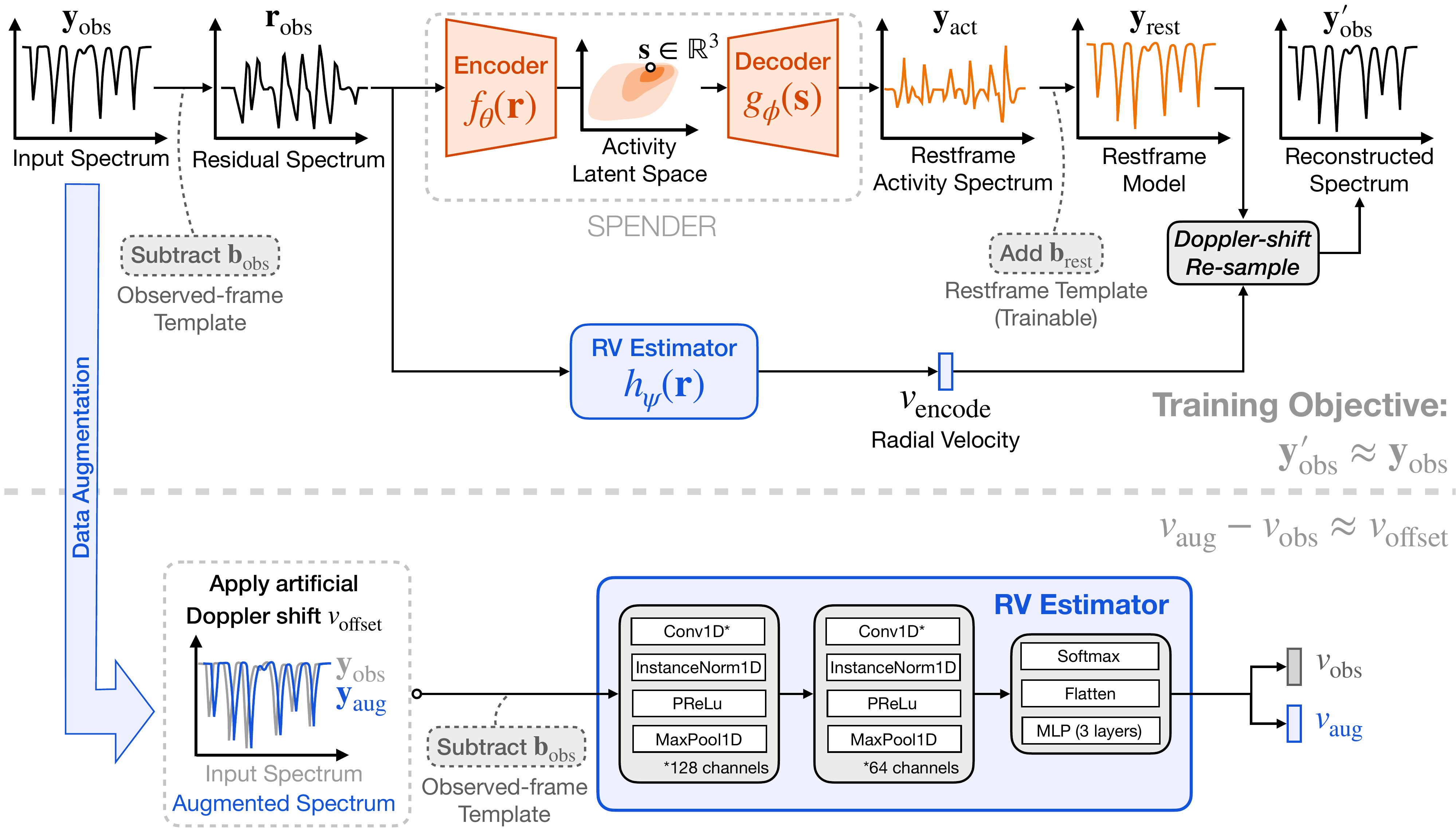}
\caption{\label{fig:diagram} 
\aestra combines a convolutional RV estimator, whose output is denoted $v_{\rm encode}$, and an attentive auto-encoder that represents activity-induced spectral features with a latent vector $\mathbf{s}$.
The decoder network expands $\mathbf{s}$ into a rest-frame spectrum $\mathbf{y}_{\rm act}$, which is shifted into the observed frame to produce the reconstructed spectrum $\mathbf{y}_{\rm obs}'$. The architecture is trained with a fidelity loss function that
ensures $\mathbf{y}_{\rm obs} \approx \mathbf{y}_{\rm obs}'$ and an RV-consistency loss function that seeks to recover 
$v_{\rm offset}$ from an artificially Doppler-shifted ``augment'' spectrum $\mathbf{y}_{\rm aug}$. 
The combined loss function is designed to disentangle activity-induced spectral features and Doppler shift-induced features.
}
\end{figure*}


\autoref{fig:diagram} illustrates the architecture of \aestra.
The input consists of a collection of hundreds or more of spectra of a single star, obtained
over a range of different activity states and different phases of the orbital motion of any planets.
The input is processed in parallel by a neutral network RV estimator and a spectrum autoencoder.
The RV estimator learns the effect of a Doppler shift: a pure stretching of the
wavelength scale, regardless of any other details of the spectrum.
\hl{
Training of the RV estimator network does not require a spectral template or line list, or indeed any prior knowledge about the star. Instead, we introduce artificial Doppler shifts into the input spectra, and optimize the RV estimator network to recover these shifts.
Within the RV estimator are filters (convolutional kernels) that sweep across the residual spectrum. Ideally, during training, the parameters of these filters are tuned to recognize the patterns that result from Doppler shifts, while also learning to ignore photon noise.
}
\hl{Thus, this part of \texttt{AESTRA} offers an alternative approach to
traditional RV extraction methods involving template matching
and cross-correlation functions.}

This RV estimator, trained to recognize pure wavelength stretching, can accurately determine velocity differences between stellar spectra exhibiting the same activity state. However, when measuring RV differences between spectra of different activity states, the RV estimator is subject to biases due to spectral distortions that mimic Doppler shifts. 
In the absence of any ``ground truth'' about the star and its motion,
we rely on a key assumption: the star's orbital motion is uncorrelated
with its intrinsic variability during the observation period. 
Assuming this condition holds, we can mitigate the activity-induced RV bias by decorrelating the RV estimator output against activity indicators. These indicators should capture the star's activity state while remaining unaffected by Doppler shifts. 

In this paper, we introduce an auto-encoder network to generate these activity indicators through spectrum modeling. The encoder compresses the information in each stellar spectrum into a small number of ``latent parameters'',
and the decoder transforms a set of latent parameters into an accurate reconstruction of a rest-frame stellar spectrum.
The architecture is designed to assign latent parameters based on spectral perturbations
that are independent of the Doppler shift of the spectrum.
In this respect, the latent parameters are generalizations
of traditional activity indicators such as the bisector span
of the CCF or the equivalent width of certain emission lines. This de-trending technique is explored in more depth in \autoref{subsec:detrend}

\subsection{Architecture}\label{subsec:architecture}

We begin with a collection of normalized one-dimensional input spectra.
Each spectrum is represented by a vector $\mathbf{y}_{\rm obs} \mathbf \in \mathbb{R}^L$,
giving the intensity at each of $L$ wavelengths $\mathbf{\lambda}_{\rm obs}$.
\hl{We do not model the Earth's barycentric motion, thereby implicitly assuming
that the barycentric corrections have been applied without error.
To prepare the input for \aestra, we first establish a template spectrum, $\mathbf{b}_{\rm obs} \in \mathbb{R}^L$ that captures the star's average spectral features. This template is subtracted from each observed spectrum to create a residual spectrum
$\mathbf{r}_{\rm obs} = (\mathbf{y}_{\rm obs} - \mathbf{b}_{\rm obs})$.
The purpose of this step is to isolate the spectral time variations
and reduce the dynamic range of the input. In our experiments we somewhat arbitrarily chose
one of the simulated spectra with the smallest applied perturbations
to be the template. The exact choice of the template
does not matter as long as it is representative of the typical
spectrum. The only effect of small changes to the template is to apply
small time-independent offsets to all of the residual spectra, which
should not interfere with training.}

\hl{After template subtraction, the residual spectra are processed through two pipelines:}
the encoder-decoder network and the RV estimator network.

The goal of the encoder-decoder network is to summarize spectral perturbations due to stellar activity using a small number of parameters. 
The residual spectrum is compressed by a parameterized encoder $f_\theta(\cdot)$ into a latent vector $\mathbf s \in \mathbb{R}^S$. Here, $\theta$ denotes the adjustable parameters in the encoder.
\begin{align}
    \mathbf s = f_\theta(\mathbf{r}_{\rm obs}).
\end{align}
Following the approach of \cite{melchior2023autoencoding}, the encoder $f_\theta(\cdot)$ consists of three convolutional layers, an attention layer, and a three-layer Multi-Layer Perceptron (MLP). 
A more in-depth description of these machine learning components can be found in \cite{smith2023astronomia}.
The attention layer is integrated to help the encoder focus on significant spectral features amidst varying positions and noise. In practice, the attention layer works by dividing the extracted feature matrix into two sub-matrices. The first represents the identified features across different wavelengths, while the second indicates the relative significance of these features (commonly called ``attention weights''). By multiplying these matrices element wise and then summing over wavelengths, we derive ``attended'' features that emphasize the most relevant spectral information.

\begin{figure}[t]
\centering
\includegraphics[trim={2em 0 0 0},width=0.48\textwidth]{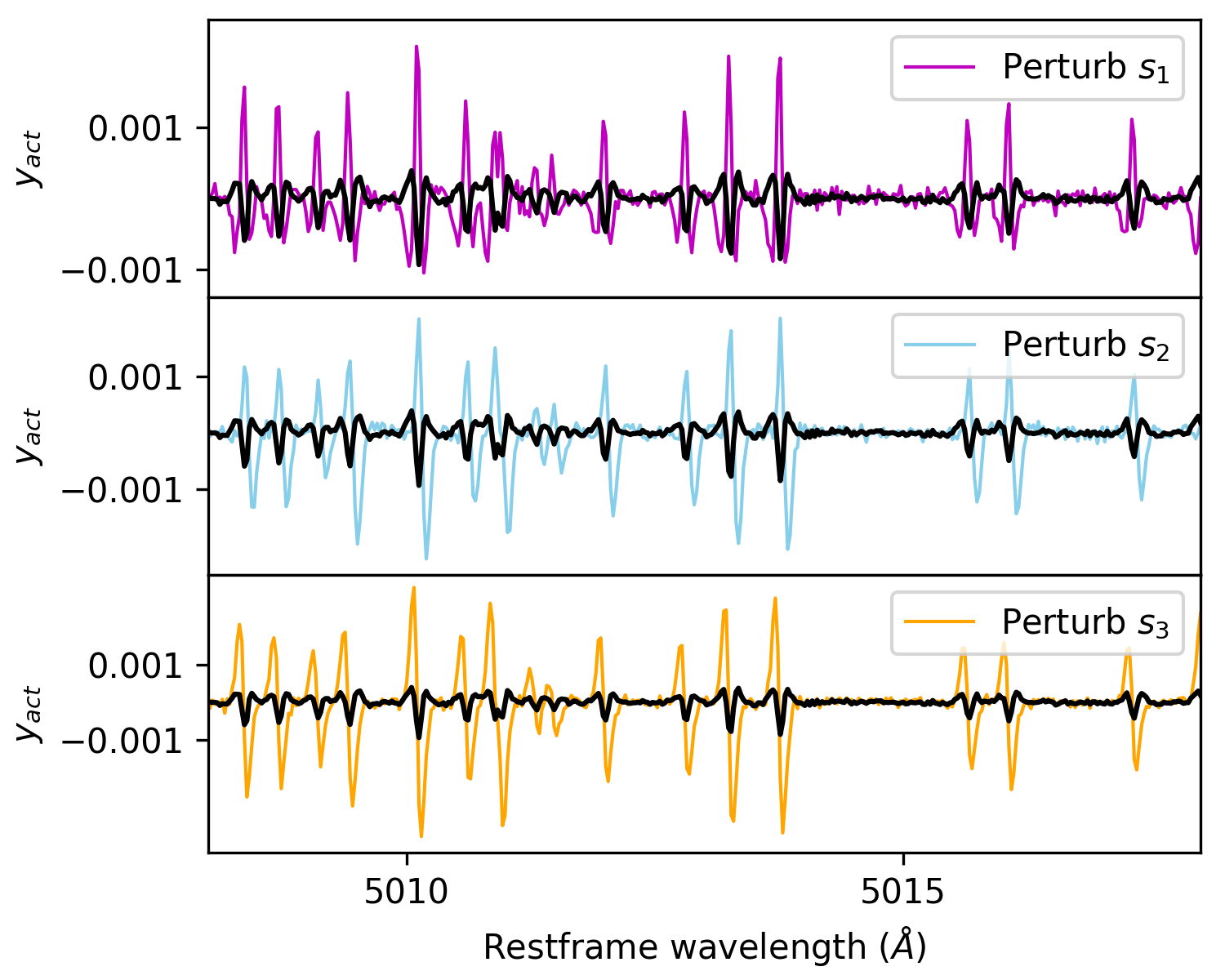}
\caption{Visualization of the effect of perturbing the latent vector based on a representative spectrum to be described in \autoref{sec:application}. The black curve shows the activity spectrum $\mathbf{y}_{\rm act}$ generated from a particular choice of the latent vector $\mathbf s$.
The colored lines show the resulting activity spectrum after perturbing each component of the latent vector: perturbing $s_1$ (purple) appears to introduce line width changes, while perturbations in $s_2$ (blue) and $s_3$ (orange) produce features related to line asymmetries.}
\label{fig:y_act}
\end{figure}

\hl{Conceptually, we may regard the components of the latent vector as generalizations
of traditional activity indicators, such as line widths or bisector spans.}
We chose a dimension $S=3$ for the latent space as a balance between model complexity, which favors large $S$, and interpretability, which favors small $S$.
This relatively low-dimensional latent space simplifies the interpretation of the latent vector as activity indicators.
\hl{As an illustration of the meaning of the latent vector, \autoref{fig:y_act} shows the effect of perturbing each component, one at a time,
using the model to be described in Section~3.3. In this case, $s_1$ appears to be
related mainly to line widths, and $s_2$ and $s_3$ to line asymmetries.}


The decoder function $g_\phi(\cdot)$ with adjustable parameters $\phi$ is another three-layer MLP which uses $\mathbf s$ to generate an activity spectrum $\mathbf{y}_{\rm act}$ in the stellar intrinsic rest frame, represented as a vector of length $L'$ with a slightly extended wavelength coverage ($L'>L$):
\begin{align}
    \mathbf{y}_{\rm act} = g_\phi(\mathbf s).
\end{align}

\noindent To encourage the decoder to focus on the activity-induced perturbations, we add a trainable rest-frame template spectrum ${\mathbf{b}}_{\rm rest} \in \mathbb{R}^{L'}$ to the decoded activity spectrum to produce the complete rest-frame model:
\begin{align}
    \mathbf{y}_{\rm rest} = \mathbf{y}_{\rm act} + \mathbf{b}_{\rm rest}. 
\end{align}
The rest-frame template, $\mathbf{b}_{\rm rest}$, represents a hypothetical spectrum with zero activity ($\mathbf{y}_{\rm act} = 0$). \hl{The rest-frame template is distinct
from the observed-frame template described earlier. One may consider $\mathbf{b}_{\rm obs}$ to be an initial guess and $\mathbf{b}_{\rm rest}$ as an optimized model for
the baseline stellar spectrum.}
%

The RV estimation network $h_\psi(\cdot)$, visualized in \autoref{fig:diagram}, is a modified version of the encoder network. 
\hl{A key difference between the encoder and the RV estimation network is that
the encoder includes an attention layer \citep{vaswani2017attention} -- which is designed to help the
model identify the most important patterns of variability irrespective of wavelength shifts --
and the the RV estimator lacks such a layer.}

\hl{The RV estimator begins with two convolution layers that apply convolutional filters to the input data that are trained to respond maximally to Doppler shifts. 
After each convolution layer, we apply
a Parametric Rectified Linear Unit (PReLU) operation \citep{he2015delving}, which introduces non-linearity to the network and helps to enhance the recognized features. 
Next, max pooling is applied, down-sampling the features by retaining only the dominant features from each segment of the residual spectrum.
This leaves us with a
$64 \times 20$ matrix representing 64 output channels and 20 wavelength segments. A softmax operation is applied in the wavelength direction. This operation helps to quantify the relative importance to different segments.
Then, the resulting matrix is flattened into a vector with 1{,}280 elements. 
These elements are passed to a MLP with (128,\,64,\,32) nodes. The MLP learns to estimate Doppler shifts based on the feature vector summarized by the convolutional layers.} The estimated RV is denoted $v_{\rm encode}$:
\begin{align}
    v_{\rm encode} =  h_\psi(\mathbf{r}_{\rm obs}).
\end{align}
Lastly, the rest frame model spectrum $\mathbf{y}_{\rm rest}$ is Doppler-shifted using the value of $v_{\rm encode}$ determined by the RV estimator and interpolated to match the observed wavelengths $\mathbf{\lambda}_{\rm obs} \in \mathbb{R}^L$ using cubic spline interpolation.
This step accounts for the motion of the star relative to the observer and enables a direct end-to-end comparison between the observed and model spectra.

\subsection{Loss Function}\label{subsec:loss}

To ensure high-quality spectral reconstruction, we employ the end-to-end fidelity loss function of \citet{melchior2023autoencoding}. This function quantifies the agreement between the input and reconstructed spectra, averaged over a batch of $N$ spectra each of dimension $L$:
\begin{align}
L_{\rm fid} &= \frac{1}{NL}  \sum_i^{N}  \mathbf{w}_{i}\odot(\mathbf{y}_{{\rm obs},i}-\mathbf{y}_{{\rm obs},i}')^2
\end{align}
Here, $\mathbf{y}_{i, {\rm obs}}$ denotes the $i$-th input spectrum, $\mathbf{y}_{i, {\rm obs}}'$ is the reconstructed spectrum, $\mathbf{w}_i$ is the corresponding weight vector.
The operation $\odot$ represents element-wise multiplication, and the squaring of the difference is also performed element wise.

To achieve consistent RV estimations, we introduce a novel loss term, $L_{\rm RV}$. To encourage the RV estimator to focus exclusively on the signatures caused by Doppler-shifting, we exploit data augmentation, a technique used in machine learning to increase the amount and diversity of training data by applying various transformations or manipulations to the original dataset. In this case, we artificially apply a Doppler shift while preserving the spectral shape.  Data augmentation allows us to generate very large samples for training the RV estimator:
\begin{equation}
\begin{split}
&L_{\rm RV} = \frac{1}{N}\sum_i^N \left[\frac{1}{ \sigma_v^2} (v_{{\rm aug},i}-v_{{\rm obs},i}-v_{{\rm offset},i})^2 \right] \label{eq:MSE} \\
&v_{{\rm obs},i} =  h_\psi(\mathbf{r}_{{\rm obs},i}) \quad
v_{{\rm aug},i} =  h_\psi(\mathbf{r}_{{\rm aug},i})
\end{split}
\end{equation}

As shown in Eq.~\ref{eq:MSE}, we use a weighted mean squared error (MSE) loss to compare the predicted RV offset between the original and augmented spectra ($v_{{\rm aug},i}-v_{{\rm obs},i}$) to the true injected offset ($v_{{\rm offset},i}$). During training, $v_{{\rm offset},i}$ is randomly drawn from a uniform distribution $v_{\rm offset} \sim \mathcal{U}(-3, 3)$ \metersec, and $\sigma_v$ represents the Cramer-Rao theoretical limiting uncertainty of a single RV measurement based only on photon noise \citep[see, e.g.][]{bouchy2001fundamental,Beatty2015}.
Because the RV estimator is trained on augmented spectra with artificial Doppler shifts, there is no need to split the data into training, validation, and test sets. During each iteration of training (commonly referred to as an ``epoch'' in the machine learning literature), a new batch of training data is generated on the fly, ensuring that the model is consistently exposed to new Doppler shifts. 

Lastly, we introduce a regularization term to constrain the decoder's flexibility and prevent over-fitting.  As both the activity spectrum and rest-frame baseline are trainable, adjusting  $\mathbf{b}_{\rm rest}$ by adding a constant can be compensated by subtracting the same constant from  $\mathbf{y}_{\rm act}$. To resolve this degeneracy between $\mathbf{y}_{\rm act}$ and $\mathbf{b}_{\rm rest}$, we define a Ridge regularization loss as follows:
\begin{align}
L_{\rm reg} &= \frac{k_{\rm reg}}{NL}  \sum_i^{N}  \frac{\mathbf{y}_{{\rm act},i}^2}{\sigma_y^2}
\end{align}
The adjustable hyper-parameters $k_{\rm reg}$ and $\sigma_y$ determine the relative weight of the regularization loss. In practice, we set $\sigma_y=0.1$ to achieve a good balance between model flexibility and trainability. 
During optimization, we cyclically adjust the weight of regularization loss  $k_{\rm reg}$, increasing from 0 to 1 over a cycle of 1000 iterations, and then setting  $k_{\rm reg}$ to 0 and restarting the cycle.

We adopted a two-phase strategy to train \aestra models. In the first phase, the RV estimator was trained independently until $L_{\rm RV}$ approached 1, while leaving the encoder-decoder network and rest-frame baseline unchanged. In this stage, the RV estimator can accurately measure the velocity offset between spectra of the same activity state, but the RV estimates are still subject to activity-induced bias.
In the second phase, we jointly optimized all components of \aestra ---the encoder-decoder network, the rest-frame baseline, and the RV estimator---using the combined loss function $L_{\rm total}$:
\begin{equation}
\label{eq:loss}
L_{\rm total}=L_{\rm fid}  + L_{\rm RV} + L_{\rm reg}
\end{equation}

In summary, $L_{\rm total}$ is designed to optimize three key aspects during training: high reconstruction quality, consistent RV estimation, and shift-invariant encoding. The fidelity loss, $L_{\rm fid}$, ensures that the reconstructed spectra closely match the input spectra. The RV loss, $L_{\rm RV}$, promotes consistency in RV estimations by training the estimator to focus on signatures induced by stretching, using augmented data. Lastly, the regularization loss, $L_{\rm reg}$, constrains the decoder's flexibility and resolves the degeneracy between the activity spectrum and rest-frame baseline, improving training efficiency.

\begin{figure}[t]
\centering
\includegraphics[width=0.45\textwidth]{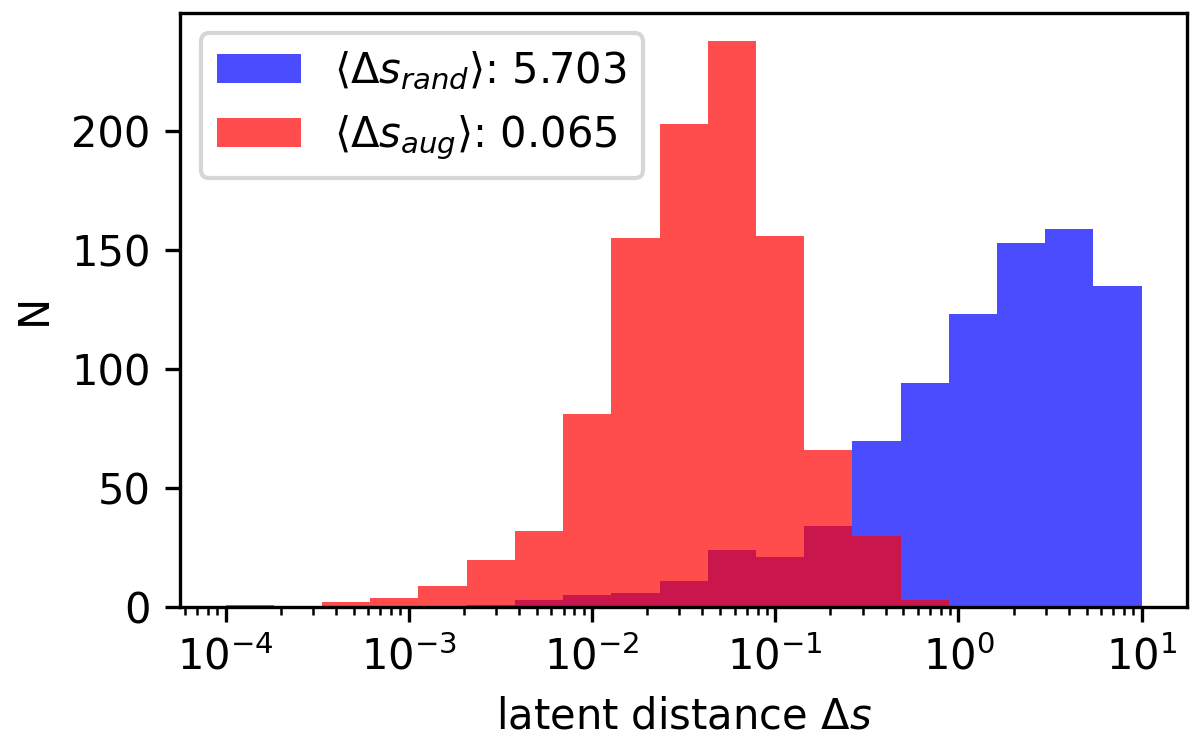}
\caption{Latent distance distribution for randomly selected data pairs ($\Delta \mathbf{s}_\mathrm{rand}$, blue), and for data-augment pairs ($\Delta \mathbf{s}_\mathrm{aug}$, red), in a dataset of 1,000 simulated spectra influenced by random activity. Augmented spectra are obtained by artificially Doppler-shifting the corresponding original spectra. For a Doppler-shift invariant encoder, the expected value of the augmented latent distances is zero: $\langle \Delta \mathbf{s}_\mathrm{aug}\rangle = 0$.
\label{fig:hist}}
\end{figure}

As an aside, we initially thought it would be necessary
to add a consistency loss term to encourage shift-invariant encoding,
following \cite{liang2023autoencoding}.
This term aims to assign the same latent vector to spectra with identical activity-induced perturbations but varying Doppler shifts by minimizing the sigmoid-modulated difference between the latent positions of the original and corresponding augmented spectra:
\begin{align}
L_{\rm c} &= \frac{1}{N}\sum_i^N {\rm sigmoid}\left[\frac{1}{\sigma_s^2 S}|\mathbf{s}_i-\mathbf{s}_{\mathrm{aug},i}|^2\right]-0.5
\end{align}
However, we found that this additional term
was not necessary for the numerical experiments we have performed.
To demonstrate, we used \aestra trained without a consistency loss term to encode augmented spectra with artificial Doppler shifts. As shown in \autoref{fig:hist}, the latent distance between the original and corresponding augmented spectra (red) exhibit a much lower dispersion than the latent distances between randomly selected data pairs (blue), indicating that the latent parameters that \aestra assigns to Doppler-shifted spectra depend only weakly on the Doppler shift.
We note this fact here because although the consistency loss term was not necessary,
it does not harm model performance and might be useful when training on real data, for
which spectral perturbations are undoubtedly more complex.

\begin{figure}[ht]
\centering
\includegraphics[width=0.45\textwidth]{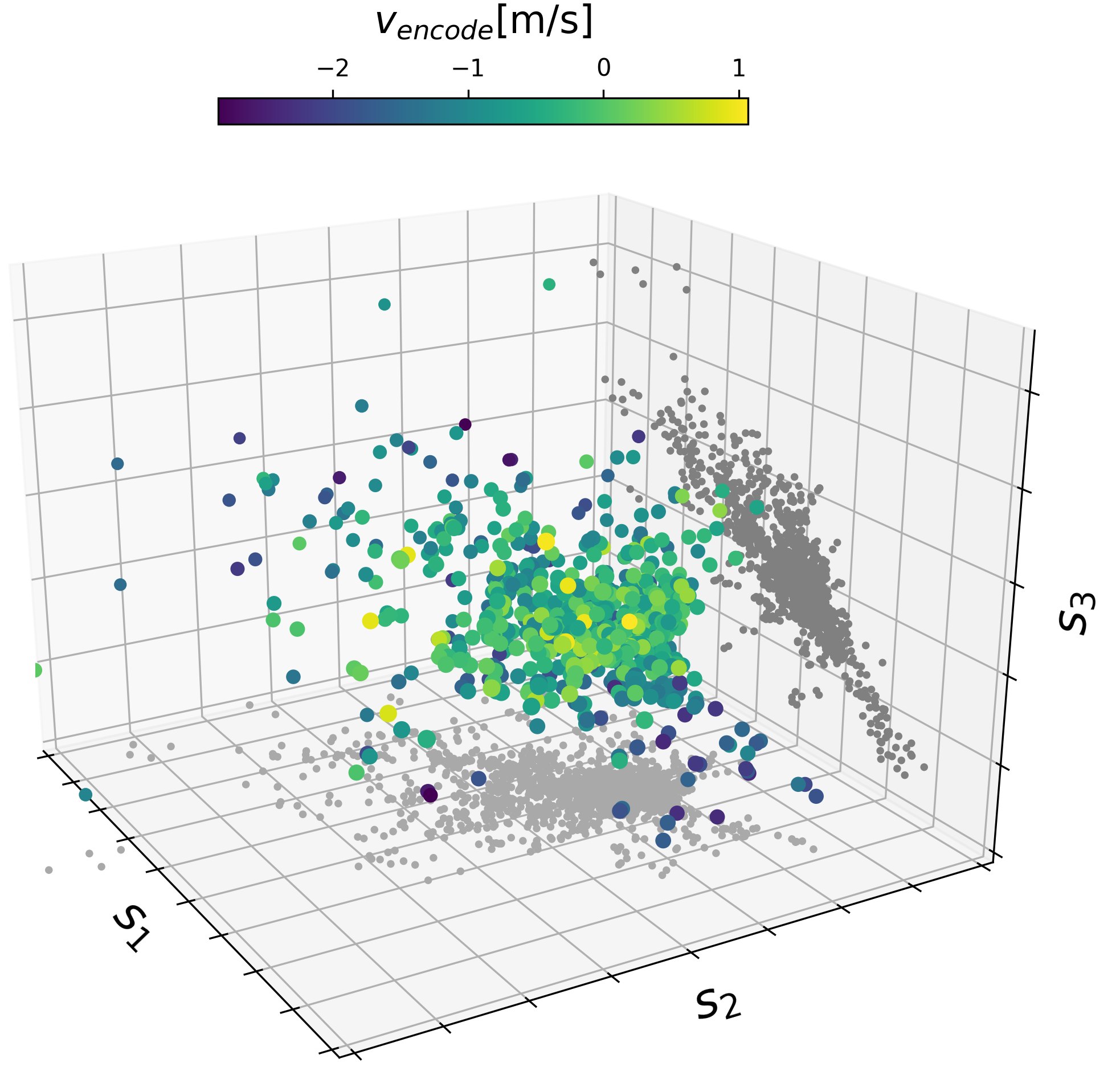}
\caption{\label{fig:3d-N1000} Distribution of 1,000 simulated spectra in latent space, color-coded according to $v_{\rm encode}$. 
A subtle color gradient from the center to the outskirts indicates a weak correlation between encoded RVs and latent position. Given that latent position can be interpreted as activity indicators, this suggests the encoded RVs are largely activity independent.
As a final correction to improve
the RV extraction, we perform Gaussian smoothing on $v_{\rm encode}$ in latent space, allowing for the extraction and subtraction of the trend.}
\end{figure}

\subsection{De-trending RVs with Latent Vectors}\label{subsec:detrend}

The star's RV estimates should ideally be independent of the spectral perturbations.
However, \autoref{fig:3d-N1000} shows that the encoded RVs are mildly correlated with latent positions.
Causal connections between activity and orbital motion
are unlikely; although hot Jupiters and other close-orbiting planets may influence a star's rotation and photospheric activity through tidal or magnetic interactions, such effects are negligible for habitable-zone exoplanets, the focus of this study.
Nevertheless, even if causal correlations are absent, Doppler shifts and stellar perturbations may be correlated because of similarities in timescales --- if, for example, the planet's orbital period is comparable to the star's rotation period or the period of a long-term magnetic activity cycle. To disentangle these effects, data should be collected in sufficient quantities and over all relevant timescales.

But we suspect another effect is responsible for the mild trend observed in \autoref{fig:3d-N1000}.
The RV estimator is trained to determine accurate RV offsets from pure spectral stretching, leaving the RV zero-points unspecified. 
The encoded RVs may be affected by stellar activity and thus contain activity-dependent RV zero-points $v_0$ that can bias our RV estimates. 
However, we have a trained activity autoencoder, capable of producing high-quality reconstructions with varying stellar activity.
We thus consider the latents $\mathbf{s}$ as generalized activity indices and now seek to determine the spurious velocity zero-points. 
We make the ansatz:
\begin{align}
    v_{{\rm encode},i} = v_{{\rm true},i} + v_0 (\mathbf{s}_i) + {\rm noise.}
\end{align}
Assuming that we have enough spectra to observe different true RVs at approximately fixed activity $\mathbf{s}$, we can isolate a constant baseline 
\begin{equation}
\begin{split}
v_{\rm 0}(\mathbf{s}_i) &\approx \langle v \rangle_{i} = \frac{\sum_{j\neq i} w_{ij} v_{{\rm encode},j}}{\sum_{j\neq i} w_{ij}},\\
{\rm where}~w_{ij} &= \exp{\left[-\frac{|\mathbf{s}_i-\mathbf{s}_j|^2}{2\sigma^2_R}\right]}.
\end{split}
\end{equation}
Here, $w_{ij}$ denotes the weight of the encoded RV $v_{{\rm encode},j}$ for neighbor $j$, and $\sigma_R$ denotes the characteristic radius of the latent distribution.
We calculate $\sigma_R$ as the average distance among the 10 nearest neighbors in the distribution. This calculation is mathematically equivalent to Gaussian smoothing in latent space. In theory, each point in the distribution should have a non-zero weight and be included in the calculation, but samples farther than 3-$\sigma_R$ exert negligible effects. In practice, the summation is carried out over the 5\% nearest neighbors in latent space, with a minimum of 10 nearest samples.

To obtain our final estimate of the true Doppler RVs, we subtract the activity-dependent RV zero-points from the encoded RVs:
\begin{align}
    v_{{\rm correct},i} = v_{{\rm encode},i} - \langle v \rangle_{i}
\end{align}
The resulting corrected RVs, $v_{{\rm correct},i}$, should be closer to the true Doppler RVs, $v_{{\rm true},i}$. We expect the efficacy of this method to increase with larger sample sizes because that we can observe the star with at different Doppler shifts in nearly the same activity state. 

\section{Applying \aestra to Simulated Data}\label{sec:application}

\subsection{Generating Synthetic Spectra from SOAP CCFs} \label{subsec:data}

To assess the performance of \aestra, we simulated a large number of sythetic spectra
spectra affected by magnetic activity.

First, we generated a ``quiet'' spectrum, $\mathbf{y}_{\rm quiet}$, without any
activity perturbations or Doppler shift. We used an evenly sampled wavelength grid ranging from 5000\AA \xspace to 5050\AA \xspace with a \hl{wavelength spacing} of 0.025\AA \footnote{
Our synthetic spectra offer better line sampling than typically seen in Sun-like stars. Additional tests with narrower line widths (0.05\AA) show that \aestra's performance is not significantly impacted by poor line sampling.} \xspace, resulting in an observed wavelength vector $\mathbf\lambda_{\rm obs}$ of length $L=2000$.

\begin{figure}[t]
\centering
\includegraphics[width=0.48\textwidth]{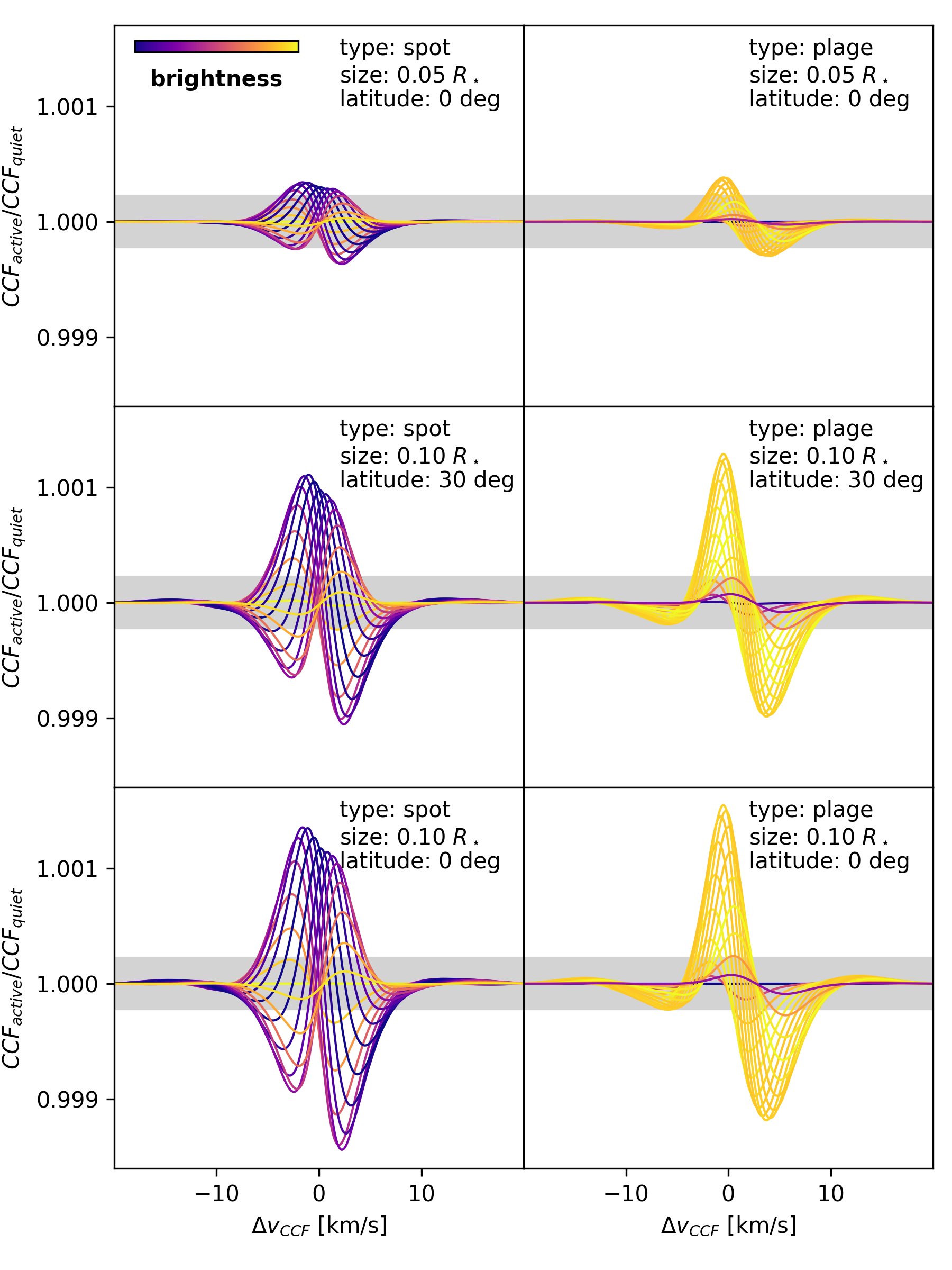}
\caption{
CCF ratios (${\rm CCF_{\rm active}/CCF_{\rm quiet}}$) versus velocity offset ($\Delta\mathbf{v}_{\rm CCF}$) for a solar-type star with a single active region, calculated by the SOAP 2.0 code. 
The color of each curve indicates the brightness variations due to the
darkening effect of the active region at different phases of stellar rotation.
The left panels depict the effect of spots with varying sizes and latitudes, and the right panels show the impact of plages. The gray shaded area represents the approximate photon noise level corresponding to a single-spectrum RV precision of 0.3 \metersec. 
\label{fig:ccf}}
\end{figure}

To produce realistic absorption lines, we randomly selected 70 line locations within the observed wavelength range. For each line, we determined a random width from a Gaussian distribution $r\sim N(\mu=0.2{\rm \AA},\sigma=0.02{\rm \AA})$ and a random amplitude from a uniform distribution $a\sim U(0.01,0.70)$.
We then calculated the quiet spectrum as the product of all absorption lines:
\begin{align}
    \mathbf{y}_{\rm quiet}=\prod_k^K \left(1-a_k\exp{\left[-\frac{1}{2} \frac{(\lambda_{\rm obs}-\lambda_k)^2}{r_k^2}\right]}\right).
\end{align}
\hl{The spectral deformations are based on the SOAP 2.0 code (Spot Oscillation And Planet; \cite{dumusque2014soap}). Since SOAP 2.0 only calculates the impact of spots and plages on the CCF rather than the entire spectrum, we needed to devise a method to perturb the synthetic spectrum using the CCFs.}

We define ${\rm CCF_{quiet}}$ as the quiet CCF of a Sun-like star evaluated at velocity offsets $\Delta\mathbf{v}_{\rm CCF}$, and ${\rm CCF_{active}}$ as the CCF affected by activity.
We introduced spectral perturbations to the simulated spectrum by perturbing all the absorption lines using the same ${\rm CCF_{active}}$.
At each epoch, ${\rm CCF_{active}}$ was calculated by randomly placing four active regions on the star's surface. The activity type (spot/plage), size, and phase of each active region were randomly drawn from the priors provided in \autoref{tab:priors}.
For each realization of ${\rm CCF_{active}}$, we defined an activity function $f_{\rm act}$ to measure the activity-induced perturbation as a function of velocity offset:
\begin{align}
f_{\rm act}(\Delta\mathbf{v}_{\rm CCF}/c) = {\rm CubicSpline} \left(\frac{\Delta\mathbf{v}_{\rm CCF}}{c},\frac{\rm CCF_{active}}{\rm CCF_{quiet}} \right).
\end{align}

\begin{figure*}[t]
\centering
\includegraphics[width=\textwidth]{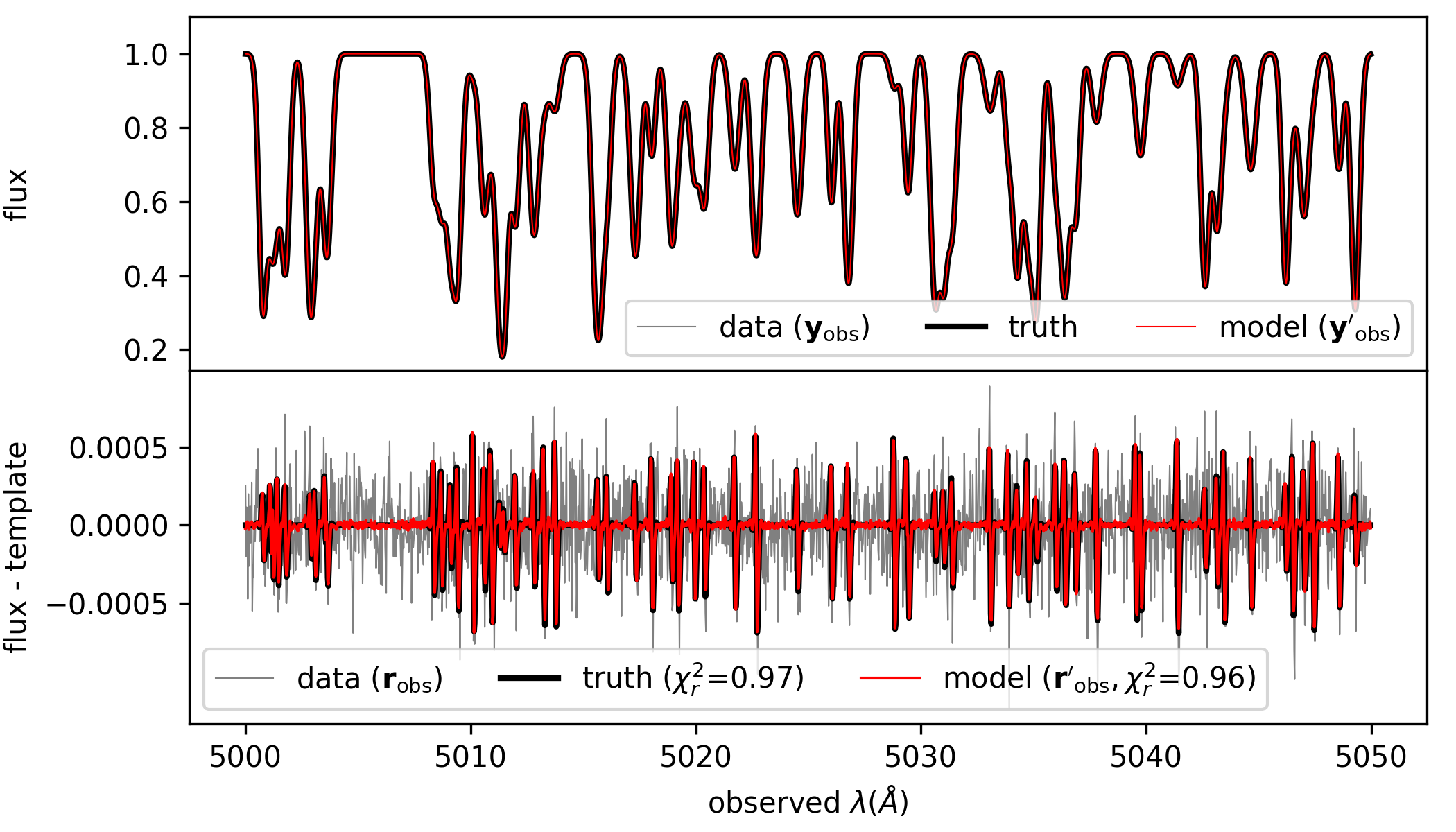}
\caption{\label{fig:spec} Simulated spectrum of a star affected by multiple active regions and photon noise ($\mathbf{y}_{\rm obs}$, gray), its reconstruction ($\mathbf{y}'_{\rm obs}$, red), and the noise-free underlying spectrum (black). In the top panel, $\mathbf{y}_{\rm obs}$
and $\mathbf{y}'_{\rm obs}$ are difficult to distinguish because
they overlap nearly perfectly. The bottom panel displays the corresponding residual spectra, with the gray spectrum predominantly showing photon noise.}

\end{figure*}

\autoref{fig:ccf} shows a collection of ${\rm CCF_{active}/CCF_{quiet}}$ due to a single active region. To perturb a specific line, we aligned the activity function $f_{\rm act}$ with the line center $\lambda_k$ and interpreted the wavelength offset as a velocity offset, given by $\Delta \lambda/\lambda_k = \Delta\mathbf{v}_{\rm CCF}/c$. By incorporating the individual line perturbations, we constructed the entire intrinsic spectrum:
\begin{align}
\mathbf{y}_{\rm intrinsic} = \mathbf{y}_{\rm quiet} \odot \prod_k^K f_{\rm act}\left( \frac{\lambda_{\rm obs}-\lambda_k}{\lambda_k} \right).
\end{align}

\begin{table}[ht]
  \centering
  \caption{Priors of the SOAP active region parameters}
    \begin{tabular}{lcc}
    \toprule
    \textbf{Parameter} & \textbf{Distribution} & \textbf{Prior} \\
    \midrule
    Activity type & Binary & $P({\rm Spot})=P({\rm Plage})=0.5$ \\
    Size [$R_\star$] & Gaussian & $\mu = 0.10$, $\sigma = 0.05$ \\
    Longitude [$^\circ$] & Uniform & $0^\circ \leq \phi < 360^\circ$ \\
    Latitude [$^\circ$] & Uniform & $-30^\circ \leq \theta \leq 30^\circ$ \\
    \bottomrule
    \end{tabular}%
  \label{tab:priors}%
\end{table}%

For the recovery test, we Doppler-shifted the perturbed intrinsic spectrum according to the true RV and interpolated it onto the grid of observed wavelengths using a cubic spline function. To simulate photon noise, we added independent Gaussian random numbers
to the spectrum:
\begin{align}
    \mathbf{y}_{\rm obs} = {\rm DopplerShift}(\mathbf{y}_{\rm intrinsic},v_{\rm true}) + {\rm noise.}
\end{align}

The true planetary RV signal at time $t$ was taken to be $v_{\rm true}(t) = K \sin{\left( {2\pi t}/{P}\right)}$, where $K$ is the velocity semiamplitude and $P$ is the orbital period.
For the following tests, we adjusted the photon-noise level of the simulated spectra
to correspond to a Cramer-Rao theoretical limit of 0.3 \metersec on the precision of a single RV measurement.  The value of 0.3 \metersec
was chosen to be comparable to the anticipated capabilities of
frontier EPRV instruments.

\subsection{Defining RV Estimates for Model Assessment}

To evaluate the performance of the \aestra model, we track four radial velocity estimates:

\textbf{Apparent RV ($v_{\rm apparent}$):} The apparent RV is computed by finding
the Doppler shift that brings the quiet baseline spectrum and the observed spectrum into the best possible agreement (minimum $\chi^2$). Affected by stellar activity, this metric provides insight into the RV noise level due to line distortions.

\hl{\textbf{De-trended RV using traditional activity indicators ($v_{\rm traditional}$):} This quantity represents the traditional
technique for activity mitigation whereby the apparent RVs are corrected to control for the observed variations in activity indicators that
are derived from the spectra.
We use a quiet spectrum ($\mathbf{y}_{\rm quiet}$) as a template to extract the CCF of the observed, activity-perturbed spectrum ($\mathbf{y}_{\rm obs}$). Three activity indicators are measured: the bisector span, the full-width at half-maximum of the CCF (FWHM), and the depth of the CCF. Following
\cite{dumusque2014soap}, the 
bisector span is defined as the difference between the top (10\%$-$40\% depth) and bottom (60\%$-$90\% depth) bisectors.}
\hl{Multilinear regression is performed between the time
series of $v_{\rm apparent}$ and the time series
of the three activity indicators, and the results are used to
subtract the estimated contributions to the RV variations due to changes in the activity indicators.}

\textbf{Corrected RV ($v_{\rm correct}$):} \hl{This quantity is derived from the \texttt{AESTRA} model by de-trending the output of the RV estimator ($v_{\rm encode}$) against the components of the
latent vector, i.e., the generalized activity indicators (see \autoref{subsec:detrend} for details).} These values can be used to assess the model's performance in isolating true Doppler RVs from activity-induced RV offsets.

\textbf{Reference RV ($v_{\rm ref}$):} Serving as a benchmark, the reference RV is calculated by fitting the noise-free underlying rest-frame spectrum ($\mathbf{y}_{\rm intrinsic}$) to the observed spectrum ($\mathbf{y}_{\rm obs}$). The scatter in the reference RVs around the true RVs is slightly higher than the Cramer-Rao bound of
0.30 \metersec, probably due to interpolation imperfections. This serves as a reference for the best possible RV precision in the absence of stellar activity.

For the $i$-th input spectrum, let $v_i$ denote the RV measurement and $v_{{\rm true},i}$ denote the true planetary RV. We define RV scatter as the standard deviation of the residual RV ($\Delta v_i = v_i - v_{{\rm true},i}$):
\begin{align}
    (\Delta v)_{\rm std} \equiv \sqrt{ \frac{1}{N_{\rm spec}} \sum_i (\Delta v_i - ({\Delta  v})_{\rm avg})^2 }
\end{align}

\noindent \hl{To assess the model's performance, we compare $v_{\rm correct}$ to both $v_{\rm traditional}$ and $v_{\rm ref}$. An effective model should yield values for $v_{\rm correct}$ that approach the benchmark values of $v_{\rm ref}$, and an advantageous model should
reduce the RV scatter below the scatter in $v_{\rm traditional}$.}

\subsection{Case I: Time-Independent Activity}

Our overall goal was to see if \aestra could recover 0.10~\metersec Earth-like signals in the presence of RV jitter that is approximately 30 times larger in amplitude. 
First, we considered spectral perturbations that are randomly and independently assigned to each spectrum.
We generated 1,000 synthetic spectra using SOAP CCFs, with each spectrum affected by four active regions randomly drawn from the priors in \autoref{tab:priors}, without any temporal structure imposed. The apparent RVs, derived by fitting a quiet baseline to the simulated spectra, are predominantly influenced by random stellar activity.

We trained an \aestra model using a two-phase strategy as described in \autoref{subsec:loss}.\footnote{For this case and for the others described in this paper, training required several hours on an NVIDIA A100 GPU.} 
\autoref{tab:loss} summarizes the training results, and \autoref{fig:spec} illustrates an example spectrum and its reconstruction. The \aestra model achieved high reconstruction quality and RV consistency, with $L_{\rm fid} \approx 1$ and $L_{\rm RV}<1$.  
The fact that the self-reported RV precision was better than the photon noise limit suggests that the RV estimator performed well and was not the limiting factor in the RV determinations. Instead, the achievable RV precision was mainly limited by the decorrelation process.


\begin{table}[t]
  \centering
  \caption{Model performance on simulated datasets with random stellar activity (Case I) and starspot evolution (Case II). $L_{\rm reg}$ is evaluated assuming $k_{\rm reg}=1$.}
    \begin{tabular}{lccccc}
    \toprule
    \textbf{Model} & $\mathbf{N_{spec}}$ & $\mathbf K$ [\metersec] & $\mathbf{L_{fid}}$ & $\mathbf{L_{RV}}$ & $\mathbf{L_{reg}}$ \\
    \midrule
    Case I & 1000 & 0.10 & 1.007 & 0.587 & 0.025\\
    Case I & 100 & 0.30 & 0.973 & 0.559 & 0.034\\
    Case II & 200 & 0.30 & 0.982 &  0.604 & 0.025\\
    \bottomrule
    \end{tabular}%
  \label{tab:loss}%
\end{table}%

\begin{figure}[ht]
\centering
\includegraphics[width=0.45\textwidth]{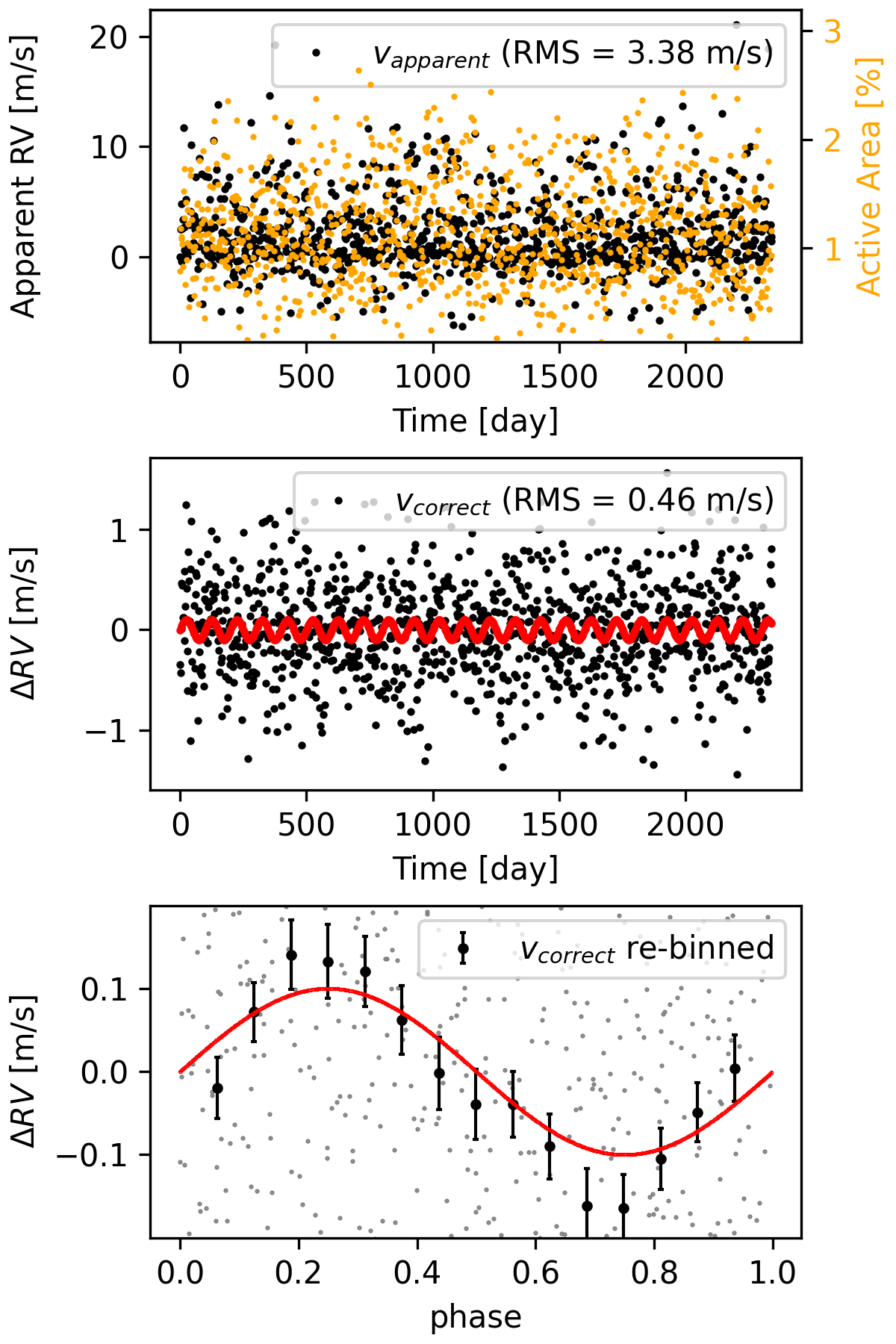}
\caption{\label{fig:N1000-rv}\textit{Top:} Time series of apparent RV (black, left y-axis) and the fraction of active area on the stellar surface (\hl{orange}, right y-axis), dominated by random stellar activity. 
\textit{Middle:} The corrected model RVs (black) are compared to the 0.1 \metersec true planetary signal (red).
\textit{Bottom:} The phase-folded and re-binned model RVs (black)  are in good agreement with the true planetary signals (red). 
}
\end{figure}

\begin{figure}[ht]
\centering
\includegraphics[trim={2em 0 0 0},width=0.48\textwidth]{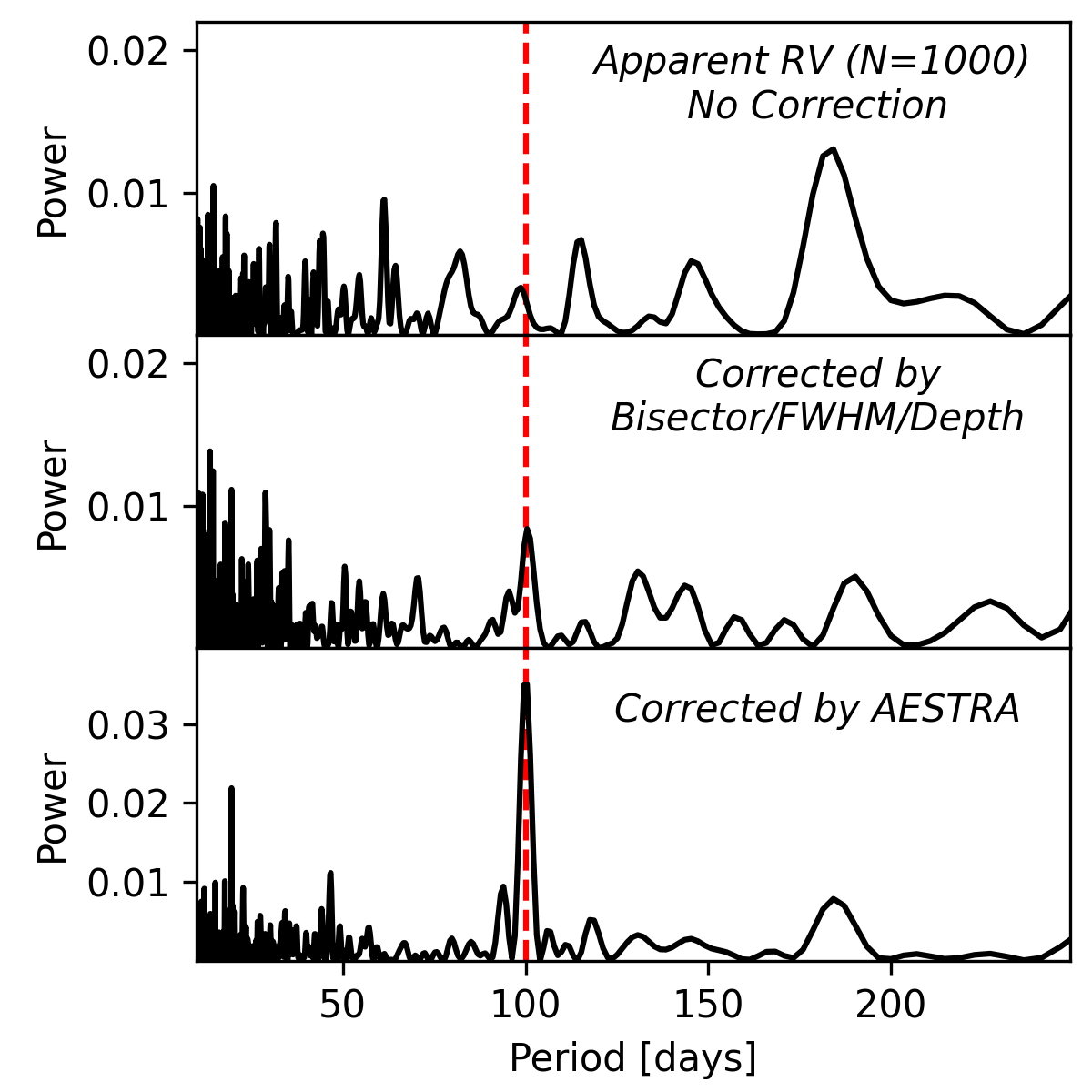}
\caption{\label{fig:period-N1000} \hl{\textit{Top:} Lomb-Scargle periodogram of the apparent RVs ($v_{\rm apparent}$, without any correction) affected by four random active regions (spots and plages).
\textit{Middle:} De-trended RVs using traditional indicators ($v_{\rm traditional}$) reveal a subtle peak at the true period (red dashed line).
\textit{Bottom:} \texttt{AESTRA} corrected RVs ($v_{\rm correct}$) show a prominent peak at the correct period}.}
\end{figure}

\begin{figure}[htb]
\centering
\includegraphics[width=0.45\textwidth]{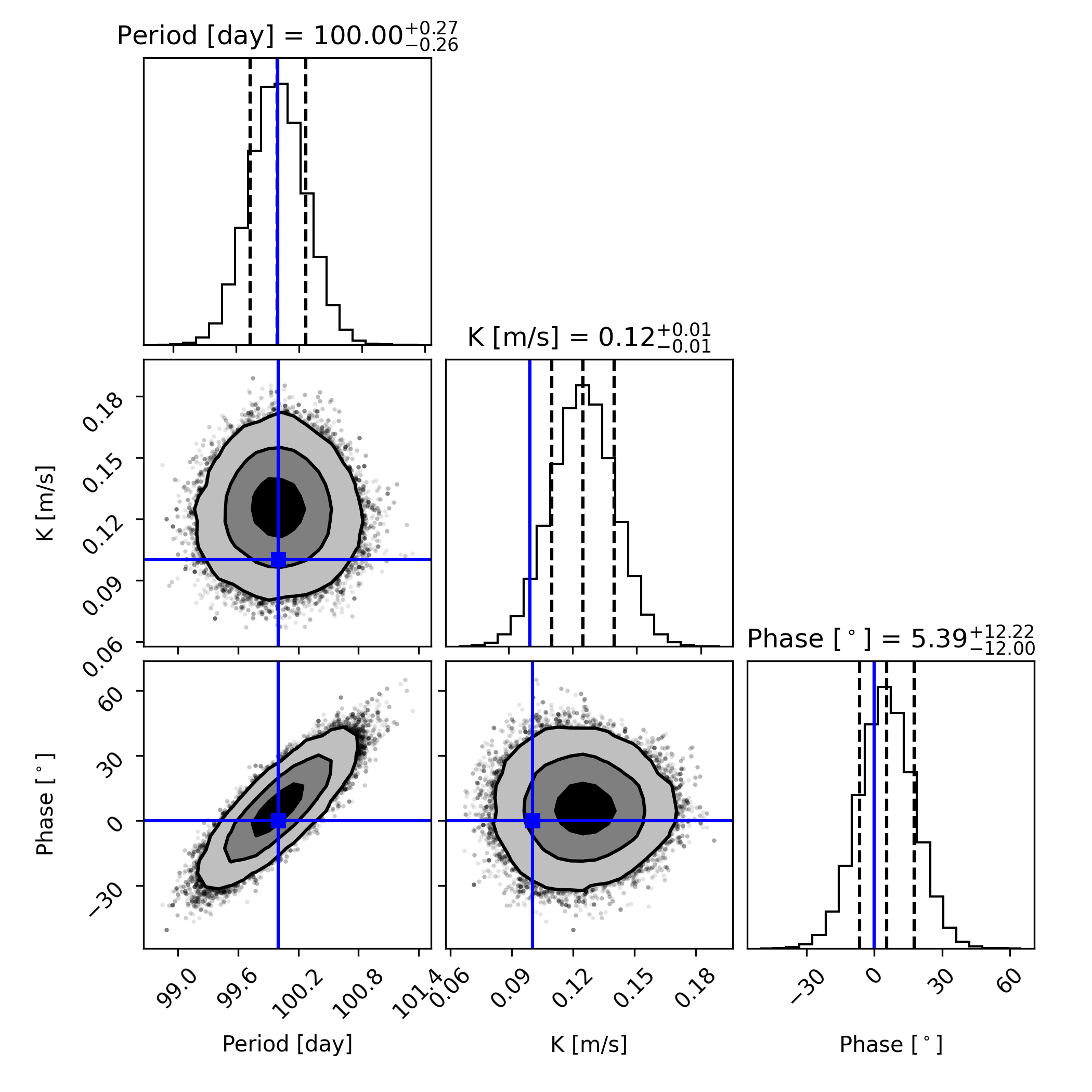}
\caption{Corner plot illustrating the posterior distribution of the orbital parameters for 1{,}000 synthetic spectra affected by random activity, with truth values indicated by blue lines. 
\label{fig:corner}
}
\end{figure}

\hl{\autoref{tab:results} summarizes the performance of different methods for extracting RVs from the simulated dataset of 1,000 spectra.
The method involving traditional activity indicators managed to reduce the apparent RV scatter from 3.38 \metersec to 0.98 \metersec.
The \texttt{AESTRA} model achieved a further reduction of the RV scatter to 0.46 \metersec, closely approaching the RV benchmark for this dataset. 
\autoref{fig:period-N1000} presents Lomb-Scargle periodograms of the RVs de-trended by traditional activity indicators ($v_{\rm traditional}$) and those corrected by \texttt{AESTRA} ($v_{\rm correct}$).
While the traditional analysis yields a peak at the true period, there
are also spurious peaks with similar amplitudes.
The periodogram of the \aestra-corrected RVs shows a more prominent peak, allowing for a clearer detection of 0.10\,\metersec planetary signal.}

To further evaluate the model's performance in recovering planetary signals, we performed Markov Chain Monte Carlo (MCMC) sampling using the \texttt{emcee} package \citep{foreman2013emcee} to retrieve the orbital parameters from the corrected RV $v_{\rm correct}$. 
Currently, we set the uncertainty for model RVs to be the same as the RV scatter $(\Delta v_{\rm correct})_{\rm std}$, as obtaining principled uncertainties directly from \aestra remains a task for the future. We intend to explore alternative approaches, such as noise propagation through the network or least-squares fit error, in the future development of \aestra.

\begin{table}
\begin{center}
\caption{RV estimates and parameter constraints through MCMC sampling}\label{tab:results}
\setlength\doublerulesep{0.2pt}
\begin{tabular}{cccccc}
\toprule
&\multicolumn{2}{c}{\textbf{Case I}}&\textbf{Case II}\\\cmidrule{2-3}
 \textbf{RV [\metersec]}& $\mathbf{N=1000}$ & $\mathbf{N=100}$ & $\mathbf{N=200}$ \\
 \midrule
$(\Delta v_{\rm apparent})_{\rm std}$ & 3.38 & 3.27 & 3.51\\
$(\Delta v_{\rm traditional})_{\rm std}$ & 0.98 & 0.96 & 0.98\\
$(\Delta v_{\rm correct})_{\rm std}$ & 0.46 & 0.73 & 0.44\\
\bottomrule
\addlinespace\addlinespace
\multicolumn{1}{c}{\textbf{Reference RV }}&&&\\
\midrule
$(\Delta v_{\rm ref})_{\rm std}$ & 0.39 & 0.38 & 0.41\\
\bottomrule
\addlinespace\addlinespace
\multicolumn{1}{c}{\textbf{Parameters}}&&&\\
\midrule
Period [day] &${100.0}^{+0.27}_{-0.26}$ & ${115.1}^{+7.6}_{-7.8}$ & ${101.9}^{+1.3}_{-1.3}$\\
$K$ [\metersec] & ${0.12}^{+0.01}_{-0.01}$ & ${0.36}^{+0.08}_{-0.07}$ & ${0.29}^{+0.03}_{-0.03}$\\
Phase [$^\circ$] & ${5.0}^{+12.0}_{-12.0}$ & ${70.0}^{+25.0}_{-27.0}$ & ${37.0}^{+13.0}_{-13.0}$\\
\bottomrule
\end{tabular}
\end{center}
\end{table}

Assuming circular orbits for simplicity, we assigned uniform priors for the orbital period between 10 and 200 days, semi-amplitude between 0.0 and 10.0 \metersec, and phase between $-180$ and 180 degrees. We initialized 32 walkers near the maximum a posteriori (MAP) solution and ran the MCMC sampler for 10,000 steps with a 2,000-step burn-in.  As shown in \autoref{fig:corner}, the RVs measured by \aestra constrained the orbital period to be ${100.0}\pm{0.27}$ days and the signal amplitude to be ${0.12}\pm 0.01$ \metersec.  The full parameter constraints are summarized in \autoref{tab:results}. 

\begin{figure}[t]
\centering
\includegraphics[width=0.42\textwidth]{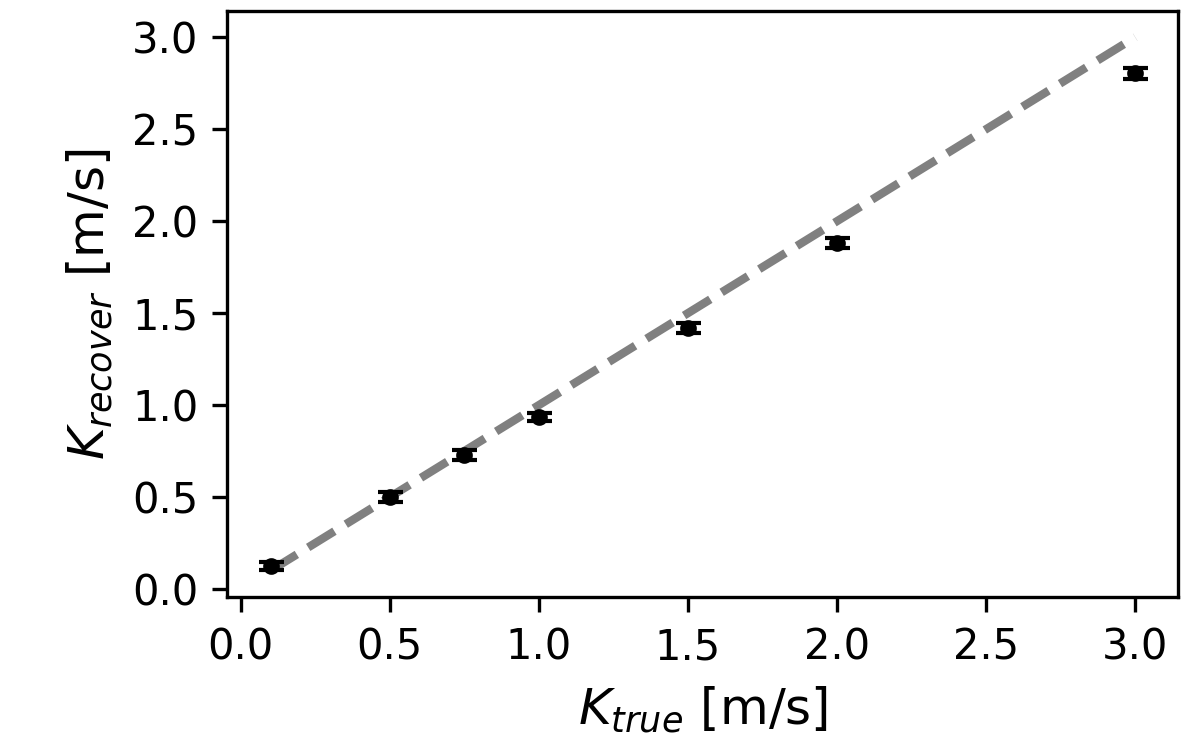}
\caption{Recovered versus true planetary signal amplitudes.  The model accurately recovers small amplitude planetary signals but tends to underestimate amplitudes $>$1 \metersec.
\label{fig:semi-amplitude}
}
\end{figure}

To assess \aestra's performance in recovering planetary signals with different amplitudes, we created new datasets using the same activity perturbations but varying injected signal amplitudes ($K$ values). The model, trained on the original dataset with 0.1 \metersec planetary signals, was applied to recover the signal amplitudes in the new datasets. 
\autoref{fig:semi-amplitude} shows that the model accurately recovered weaker signals and slightly underestimated amplitudes $>$ 1 \metersec. This probably occurs because planetary RVs are treated as noise during de-trending, causing stronger signals to introduce more uncertainty in activity-dependent RVs.

Deep learning methods generally require large datasets, making it important to understand how \aestra performance scales with sample size.
We generated a smaller dataset of 100 spectra affected by random stellar activity and injected a planetary signal of 0.30 m/s, comparable to the RV scatter caused by pure photon noise. The apparent RV scatter in the $N_{\rm spec} = 100$ dataset was 3.27 m/s.
We followed the same procedure to train the \aestra model. The training performance and resulting RV estimates are detailed in \autoref{tab:loss}  and \autoref{tab:results}. 

For the 100-spectrum simulated dataset, the best-fit \aestra model reduced the RV scatter from 3.27 \metersec to 0.73 \metersec\hl{, lower than
the scatter of 0.96 \metersec achieved by the method
involving traditional activity indicators}. While the improvement was not as good
as it was in the 1{,}000-spectrum dataset, the \aestra-corrected RVs still allowed for a 5-$\sigma$ detection of the injected planetary signal, and gave a correct estimate of the RV semi-amplitude,
$K={0.36}^{+0.08}_{-0.07}$ \metersec. 

We believe that for small sample sizes, the primary source of RV uncertainty in $v_{\rm correct}$ arises from the decorrelation process, which relies on Gaussian smoothing in the latent space to estimate the activity-dependent RV zero-points. If the latent space is underpopulated, i.e., if a given type of stellar perturbation is not observed at different Doppler shifts, it becomes challenging to calibrate the RV zero-points. This highlights the importance of capturing potential stellar activity perturbations through repeated observations.

\subsection{Case II: Time-Dependent Activity}

\begin{figure}[htb]
\centering
\includegraphics[width=0.45\textwidth]{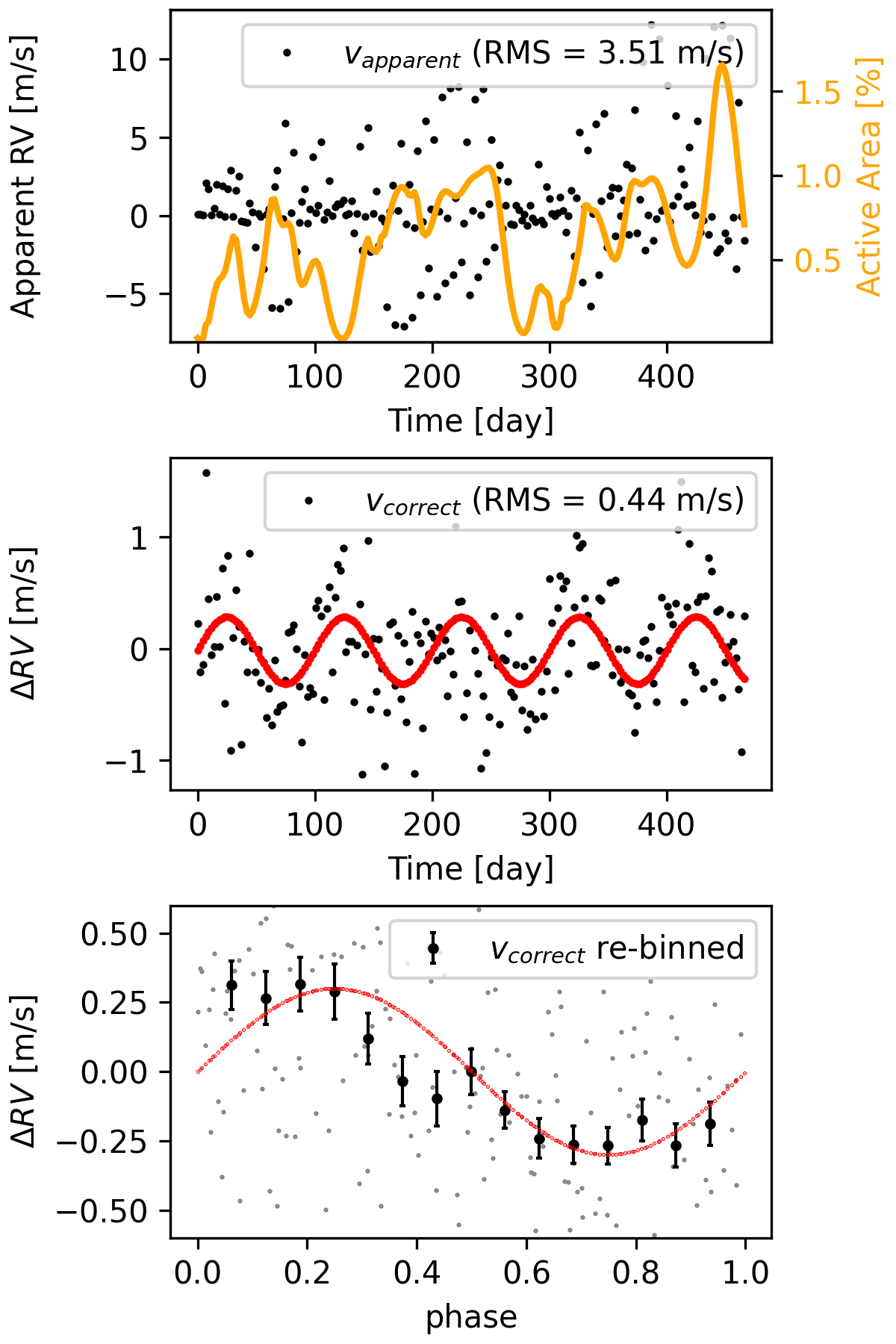}
\caption{\label{fig:N200-rv} Top panel: Time-series of apparent RVs (black, left y-axis) and the percentage of the stellar surface covered by active regions (\hl{orange}, right y-axis). Middle panel: Corrected model RVs (black) compared to injected Doppler RVs (red). Bottom panel: Phase-folded, re-binned corrected model RVs (black error bars), showing good agreement with the true RVs (red).
}
\end{figure}

\begin{figure}[h]
\centering
\includegraphics[trim={2em 0 0 0},width=0.48\textwidth]{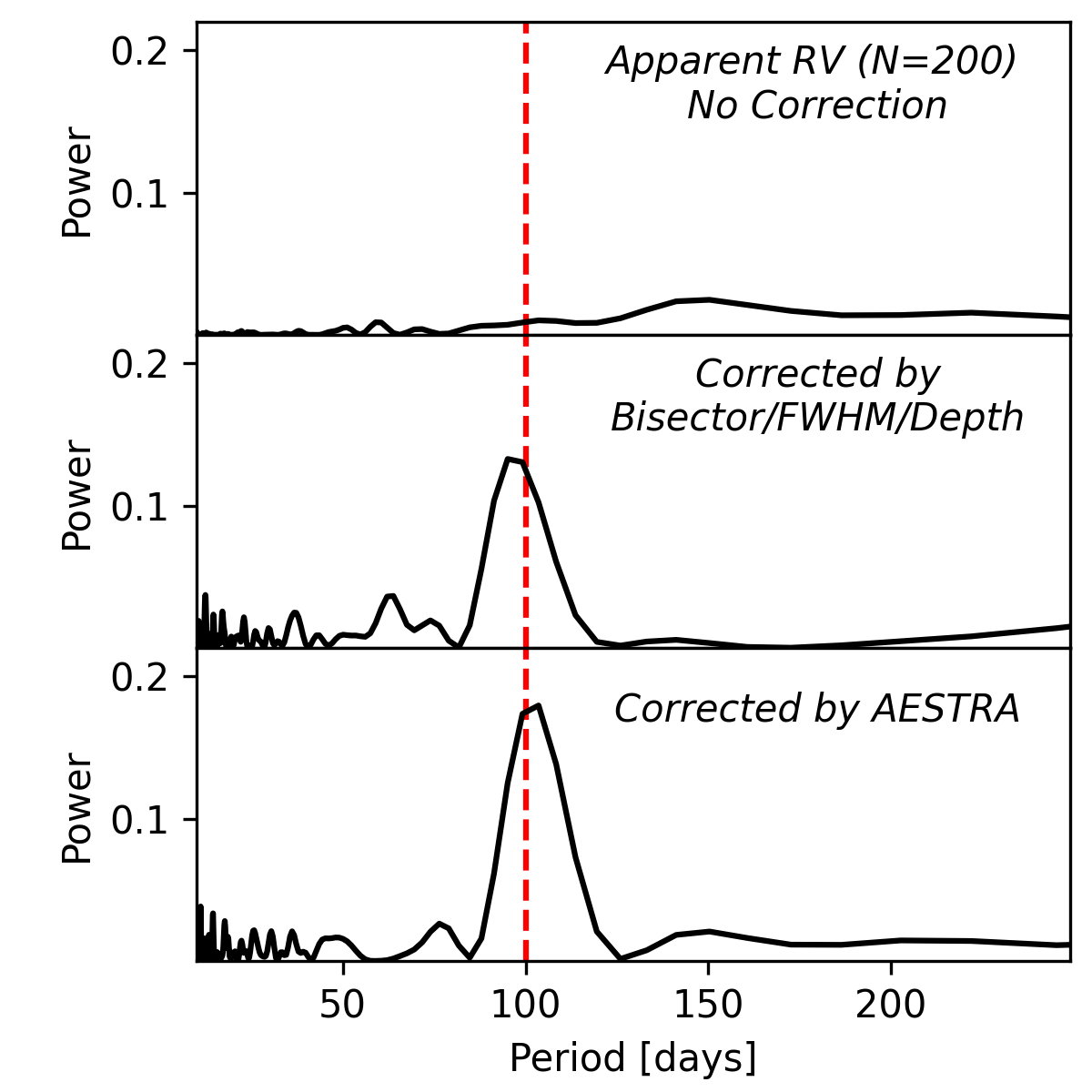}
\caption{\label{fig:period-N200} \hl{Lomb-Scargle Periodograms illustrating the impact of complex spot evolution on apparent RVs (top), RVs de-trended using traditional activity indicators (middle), and the \aestra-corrected RVs (bottom). The true planetary period (red) is detected by both methods.}}
\end{figure}

Next, we performed simulations in which stellar activity exhibited realistic temporal correlations.
We generated a dataset of 200 synthetic spectra affected by evolving starspots using the SOAP 2.0 code, incorporating the time-structured nature of starspot growth, decay, and rotation, as detailed below.
A planetary signal with
amplitude 0.30 \metersec was injected into the dataset for recovery testing.

As before, we placed four spots on the stellar surface, with new spots generated at random locations as old spots decayed. Each spot's size evolved over time, following a Gaussian function:
\begin{align}
S(t) = S_{\max} \exp\left[-\frac{1}{2}\left(\frac{t - t_{\rm peak}}{\tau}\right)^2\right],
\end{align}
where $S(t)$ represents the spot's size at time $t$, $S_{\text{max}}$ is the maximum spot size, chosen randomly between 0.5\% and 20\% stellar radius, $t_{\text{peak}}$ is the time when the spot reaches its maximum size, and $\tau$ is the spot's lifetime, determined by $\tau \propto S_{\rm max}^2$, with $\tau_{\rm max} \sim P_{\rm rot}$ for the largest spots. The spots' longitudes and latitudes were randomly chosen based on the priors described in \autoref{tab:priors}. The growth and decay of starspots results in distinct time structures in the apparent RVs, as shown in the top panel of \autoref{fig:N200-rv}.

Training the \aestra model on this dataset led to a reduction in the apparent RV scatter from 3.51 \metersec to 0.44 \metersec, as presented in \autoref{tab:results}. \hl{The \aestra-based RV scatter of 0.44 \metersec is about one-half of the scatter obtained with the
traditional method, and leads to a solid detection of the 0.3 \metersec planetary signals in the phase-folded RVs shown in \autoref{fig:N200-rv}.}

\hl{Both the \aestra\ model and the traditional analysis identified significant peaks at the true planetary period in their Lomb-Scargle periodograms (\autoref{fig:period-N200}). However, the traditional analysis gave $K = 0.52 \pm 0.09$\,\metersec, an overestimate of the true RV semi-amplitude. The \aestra\ model gave
a correct estimate, $K$ = ${0.29}\pm0.03$ \metersec.}

\hl{As summarized in \autoref{tab:results}, when the \aestra\ RVs
were fitted with a circular-orbit model using an MCMC method, 
the posteriors for the orbital parameters overlapped with
the true values, despite
the presence of complex and time-structured stellar activity. 
These results also underscore the importance of good temporal coverage for robust RV estimation.}



\section{Conclusions and Outlook}\label{sec:conclude}

\aestra is a deep learning method designed to separate Doppler shifts from spectral perturbations
due to stellar activity. It operates on a collection of spectra of the same star rather than
a single spectrum or a cross-correlation function. 
The autoencoder network summarizes spectral perturbations using a small
number of latent parameters, yielding a compact representation that can be utilized as generalized activity indices. 
The RV estimator network is jointly trained with the autoencoder to determine RV offsets. Training is performed using a combined loss function that balances the various objectives, resulting in a model that provides rest-frame spectral reconstruction and RV estimation.

In our simulations, \aestra successfully detected planetary signals with
amplitudes as low as 0.10 \metersec despite activity-induced perturbations
with amplitudes of 3~\metersec and photon noise of 0.30~m~s$^{-1}$. Additionally, \aestra showed encouraging performance even on datasets
consisting of as few as 100 spectra.
The model recovered planetary signals in a series of simulated spectra dominated by time-dependent stellar activity, even without utilizing time-domain information.

While \aestra has shown promising results on synthetic spectra,
there is no guarantee that it will be effective on real data. We plan to apply \aestra to analyze solar spectra obtained with instruments such as HARPS-N \citep{dumusque2015harps}, NEID \citep{lin2022observing}, and EXPRES \citep{Petersburg+2020_EXPRES}. 
This validation step will assess the model's real-world applicability
and will undoubtedly reveal limitations or biases that are not apparent in synthetic data. This may provide valuable insights into solar variability across different time scales and set the stage for applications to other stars, with and without known planets.

\aestra is at an early stage of development. We intend to explore some potential extensions to improve the model's performance. Currently, \aestra treats each spectrum independently, not considering the times of observation. In our tests, \aestra performed well in the presence of stellar activity perturbations that were correlated in time (Case II), but we expect that it would be even better if \aestra took time-domain information into account. The model could learn the temporal behavior of long-lived starspots, differential rotation, and acoustic oscillations, which exhibit distinct timescales. Future iterations of the \aestra model could benefit from integrating a continuous latent space time-series model and applying physically-motivated constraints on the timescales associated with different types of activity.

More broadly, the current version of \aestra is a general model that
does not incorporate any physical knowledge about stellar atmospheres or even any built-in concept of a spectral absorption line. It seems likely that integrating known physical processes and relationships that affect stellar spectra into \aestra will help it cope with more challenging and realistic datasets. For example, future iterations could make use of models for telluric absorption lines and for variations in the point-spread-function of the spectrograph. 


If \aestra can be successfully validated using solar data, perhaps after incorporating some of the improvements described above, the stage will be set for applications to exoplanet data. We will explore \aestra's potential in improving exoplanet characterization, analyzing low signal-to-noise candidates, and searching for Earth-like planets.
As we explore \aestra's potential in our future work, the code is made publicly available on GitHub (\url{https://github.com/yanliang-astro/aestra}), including a frozen version tagged ``paper-I" for reproducing the results of this work.

\begin{acknowledgments}
We thank Andrew Howard and David Hogg for constructive feedback and support. 
\end{acknowledgments}

%


\software{\texttt{spender} \citep{melchior2023autoencoding},  
          \texttt{SOAP} \citep{dumusque2014soap},
          \texttt{emcee} \citep{foreman2013emcee},
          \texttt{astropy} \citep{robitaille2013astropy}
          }





\bibliography{main}{}
\bibliographystyle{aasjournal}



\end{document}